\documentclass[twocolumn,aps,prc,showpacs,floatfix]{revtex4}
\usepackage{graphicx}
\usepackage{dcolumn}
\usepackage{bm}

\begin{document}

\title{Constraining nucleon high momentum in nuclei}

\author{Gao-Chan Yong}

\affiliation{%
{Institute of Modern Physics, Chinese Academy of Sciences, Lanzhou
730000, China}
}%

\begin{abstract}
Recent studies at Jefferson Lab show that there are a certain proportion of
nucleons in nuclei have momenta greater than the so-called nuclear Fermi
momentum $p_{F}$. Based on the
transport model of nucleus-nucleus collisions at intermediate
energies, nucleon high momentum caused by the neutron-proton short-range
correlations in nuclei is constrained by comparing with $\pi$ and photon
experimental data and considering some uncertainties. The high momentum cutoff value $p_{max}$ $\leq$ 2$p_{F}$
is obtained.

\end{abstract}

 \maketitle


The picture of nucleons have maximal momentum --- so called the
Fermi momentum $p_{F}$ --- in a nuclear system and roughly move
independently in the mean field created by their mutually
attractive interactions has been established since the 1950s.
However, recent proton-removal experiments using electron beams
with energies of several hundred MeV showed that only about 80\%
nucleons participate in this type of independent particle motion
\cite{e93,e96,sci08}. And high-momentum transfer measurements
have shown that nucleons in nuclei can form pairs with larger
relative momenta and smaller center-of-mass momenta
\cite{pia06,sh07}. This is interpreted by
the nucleon-nucleon tensor interaction in short range \cite{tenf05,tenf07}.
The nucleon-nucleon short-range correlations (SRC) in nuclei leads to a high-momentum
tail in single-nucleon momentum distribution above 300 MeV/c
\cite{bethe71,anto88,Rios09,yin13,Claudio15}.
And interestingly, the
high-momentum tail's shape caused by two-nucleon SRC is almost
identical for all nuclei from deuteron to very heavier nuclei
\cite{Ciofi96,Fantoni84,Pieper92,egiyan03}, i.e., roughly exhibits
a $C/k^{4}$ tail \cite{hen14,sci14,henprc15,liba15}. Nucleon
momentum distributions at even higher momenta are due to three or
many-nucleon correlations. This part of momentum-distribution
probability was deduced to be less than 1\% \cite{Egiyan2006}. We
thus in this study neglect this kind of high-momentum nucleons caused by
many-nucleon short-range correlations.

In the high-momentum tail (HMT) of nucleon momentum distribution,
nucleon component is strongly isospin-dependent, i.e., the
number of n-p SRC pairs is about 18 times that of the p-p or n-n
SRC pairs \cite{sci08}, thus in neutron-rich heavy nuclei proton
has greater probability than neutron to have momenta greater
than the nuclear Fermi momentum \cite{sci14}. In neutron stars,
the number of protons only has a small proportion. The above n-p SRC in neutron
stars will cause proton average kinetic energy far greater than
neutron's \cite{Sargsian12}. And the stronger the n-p SRC is,
the larger the difference of proton and neutron average kinetic energy is seen.

Nucleon spectral function provides fundamental
information on the dynamics of nucleon in
nuclear medium.
The nuclear momentum distribution can be obtained from the
spectral function by integrating over the excitation energy \cite{Ciofi96,wangp2013}.
The high-momentum nucleons come
predominantly from the high excitation energy regime of
the spectral function.
The
high-momentum cutoff parameter $\lambda$ (= $p_{max}/p_{F}$, i.e.,
the ratio of nucleon maximal momentum over the nuclear Fermi
momentum) was first introduced by \emph{Hen et al.} as a free parameter in the Correlated
Fermi Gas model --- an analytical approximation for the momentum distribution of nucleon in symmetric nuclei and nuclear matter \cite{henprc15}, to avoid divergence when calculating nucleon average kinetic energy assuming a $C/k^{4}$ dependence for the high-momentum tail. Therefore, the implication of the value of this \emph{effective} parameter $\lambda$ is the determination of the average kinetic energy of nucleons.

The value of average kinetic energy of nucleons and the high-momentum tail of nucleon distribution in nuclei surely affect the yields of $\pi$, $K$, $\eta$ and nucleon emission in heavy-ion collisions at intermediate energies. The isospin dependence of nucleon high-momentum distribution definitely affects transport calculations of the symmetry-energy sensitive observables. More specifically, a low (high) value of average kinetic energy of nucleons causes a small (large) number of meson production in transport calculations owing to low (high) value of collision energy of nucleon pairs. Because in neutron-rich heavy nuclei protons
have greater probability than neutrons to have momenta greater than the nuclear Fermi momentum, the high-momentum tail of nucleon momentum distribution affects values of $\pi^{-}/\pi^{+}$ ratio and the difference of neutron and proton elliptic flows \cite{zhang2016}.
Values of average kinetic energy of neutrons and protons in nuclei also
strongly affect nuclear kinetic
symmetry energy \cite{henprc15,liba15}, the latter is known plays
crucial role in both nuclear physics and astrophysics
\cite{topic14}.

In fact,
one can deduce the high-momentum
cutoff parameter $\lambda$ from nucleon momentum distribution in
deuteron \cite{hen14,henprc15} or from the high-energy electron
scattering measurements \cite{Egiyan2006,sci14}. But the high-energy
electron scattering measurements mainly probe the nucleon momenta at the
surface of nuclei \cite{Ryckebusch11}.
Since the production of $\pi^{+}$ meson in nucleus-nucleus collisions at intermediate energies
is mainly from proton-proton collision \cite{stock86}, and energetic neutron-proton scattering produces
hard bremsstrahlung photon \cite{yongp1,yongp2,yongp3}, the high momentum of nucleon in projectile or target should affect $\pi^{+}$ and hard photon productions.
In this
study, we use hadronic probe $\pi^{+}$ meson and electromagnetic probe hard photon in
nucleus-nucleus collisions to probe nucleon high-momentum cutoff
value.


To obtain the high-momentum cutoff value of nucleon in nuclei by
nucleus-nucleus collisions, we use the
Boltzmann-Uehling-Uhlenbeck (BUU) transport model \cite{bertsch},
which has been very successful in studying heavy-ion collisions at
intermediate energies. The BUU transport model describes time
evolution of the single particle phase space distribution function
$f(\vec{r},\vec{p},t)$, which reads
\begin{equation}
\frac{\partial f}{\partial
t}+\nabla_{\vec{p}}E\cdot\nabla_{\vec{r}}f-\nabla_{\vec{r}}E\cdot\nabla_{\vec{p}}f=I_{c}.
\label{IBUU}
\end{equation}
The phase space distribution function $f(\vec{r},\vec{p},t)$
denotes the probability of finding a particle at time $t$ with
momentum $\vec{p}$ at position $\vec{r}$. The left-hand side of
Eq.~(\ref{IBUU}) denotes the time evolution of the particle phase
space distribution function due to its transport and mean field,
and the right-hand side collision item $I_{c}$ accounts for the
modification of phase space distribution function by elastic and
inelastic two body collisions \cite{Dan02a,bertsch,Persram02}.
$E$ is a particle's total energy, which is equal to
kinetic energy $E_{kin}$ plus its average potential energy $U$.
While the mean-field potential $U$ of single particle depends
on its position and momentum of the particle and is given
self-consistently by its phase space distribution function $f(\vec{r},\vec{p},t)$.

\begin{figure}[h!]
\centering
\includegraphics[width=0.5\textwidth]{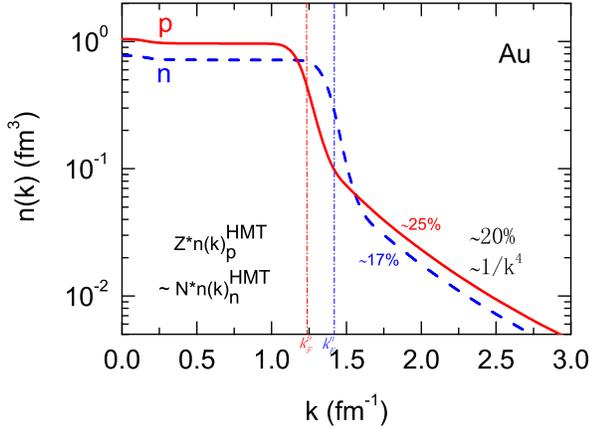}
\caption{ (Color online) Momentum distribution n(k) of neutron and proton in nucleus $^{197}_{79}Au$ with normalization condition  $4\pi\int_{0}^{\lambda k_{F}^{p,n}}n_{p,n}(k)k^{2}dk$ = 1 as well as
$4\pi\int_{k_{F}^{p}}^{\lambda k_{F}^{p}}n_{p}(k)k^{2}dk$ $\simeq$ 25\% and $4\pi\int_{k_{F}^{n}}^{\lambda k_{F}^{n}}n_{n}(k)k^{2}dk$ $\simeq$ 17\%.} \label{npdis}
\end{figure}
In the used BUU model, nucleon spatial
distribution in initial colliding nuclei is given by \cite{bertsch}
\begin{equation} \label{xyz}
r = R(x_{1})^{1/3}; cos\theta = 1-2x_{2}; \phi = 2\pi x_{3}.
\end{equation}
\begin{equation}\label{1}
    x = rsin\theta cos\phi;
    y = rsin\theta sin\phi;
    z = rcos\theta.
\end{equation}
Where $R$ is the radius of nuclei, $x_{1}, x_{2}, x_{3}$ are
three independent random numbers.
Since there is a rough 20\% depletion of nucleon momentum distribution inside the Fermi sea
\cite{sci08,yin13,sci14,Sargsian14,xu13}, we let nucleon
momentum distributions in the high-momentum tail $n^{HMT}(k) \sim 1/k^{4}$ \cite{hen14} and $\int_{k_{F}}^{\lambda k_{F}}n^{HMT}(k)k^{2}dk / \int_{0}^{\lambda k_{F}}n(k)k^{2}dk \simeq 20\%$ and keeping $n^{HMT}_{p}(k)/n^{HMT}_{n}(k) \simeq N/Z$ (N and Z being the neutron and proton numbers of a nucleus ) \cite{sci08,sci14,Sargsian14}. As shown in Fig.~\ref{npdis}, there are about 25\% (17\%) protons (neutrons) with momenta larger than the proton (neutron) Fermi momentum.
With this nucleon momentum distribution, the average kinetic energy of nucleons in this study increases roughly several MeV comparing to that with ideal Fermi-Gas model. We thus neglect this difference in heavy-ion collisions at 400 MeV/nucleon beam energy.

In this model, an isospin- and momentum-dependent mean-field
single nucleon potential is used \cite{Das03,xu14},
which reads
\begin{eqnarray}
U(\rho,\delta,\vec{p},\tau)&=&A_u(x)\frac{\rho_{\tau'}}{\rho_0}+A_l(x)\frac{\rho_{\tau}}{\rho_0}\nonumber\\
& &+B(\frac{\rho}{\rho_0})^{\sigma}(1-x\delta^2)-8x\tau\frac{B}{\sigma+1}\frac{\rho^{\sigma-1}}{\rho_0^\sigma}\delta\rho_{\tau'}\nonumber\\
& &+\frac{2C_{\tau,\tau}}{\rho_0}\int
d^3\,\vec{p^{'}}\frac{f_\tau(\vec{r},\vec{p^{'}})}{1+(\vec{p}-\vec{p^{'}})^2/\Lambda^2}\nonumber\\
& &+\frac{2C_{\tau,\tau'}}{\rho_0}\int
d^3\,\vec{p^{'}}\frac{f_{\tau'}(\vec{r},\vec{p^{'}})}{1+(\vec{p}-\vec{p^{'}})^2/\Lambda^2},
\label{buupotential}
\end{eqnarray}
where $\tau, \tau'=1/2(-1/2)$ for neutrons (protons),
$\delta=(\rho_n-\rho_p)/(\rho_n+\rho_p)$ is the isospin asymmetry,
and $\rho_n$, $\rho_p$ denote neutron and proton densities,
respectively. The parameter values $A_u(x)$ = 33.037 - 125.34$x$
MeV, $A_l(x)$ = -166.963 + 125.34$x$ MeV, B = 141.96 MeV,
$C_{\tau,\tau}$ = 18.177 MeV, $C_{\tau,\tau'}$ = -178.365 MeV, $\sigma =
1.265$, and $\Lambda = 630.24$ MeV/c are obtained by fitting
empirical constraints of the saturation density $\rho_{0}$ = 0.16
fm$^{-3}$, the binding energy $E_{0}$ = -16 MeV, the
incompressibility $K_{0}$ = 230 MeV, the isoscalar effective mass
$m_{s}^{*} = 0.7 m$, the single-particle potential
$U^{0}_{\infty}$ = 75 MeV at infinitely large nucleon momentum at
saturation density in symmetric nuclear matter, the symmetry
energy $S(\rho)$ = 30 MeV (we let kinetic symmetry energy roughly
be 0 MeV \cite{Isaac2011}) and the symmetry potential
$U^{sym}_{\infty}$ = -100 MeV at infinitely large nucleon momentum
at saturation density. $f_{\tau}(\vec{r},\vec{p})$ is
the phase-space distribution function at coordinate $\vec{r}$ and
momentum $\vec{p}$ and solved by using the test-particle method
numerically. Different symmetry energy's stiffness parameters $x$
can be used in the above single nucleon potential to mimic
different forms of the symmetry energy predicted by various
many-body theories \cite{Dieperink03} without changing any
property of the symmetric nuclear matter and the symmetry energy
at normal density.
In this study, however, both $\pi^{+}$ production in Au + Au collisions and hard photon production
in C + C collisions are in fact not sensitive to the symmetry
energy parameter $x$.

According to baryon effective mass, the isospin-dependent
baryon-baryon ($BB$) scattering cross section in medium $\sigma
_{BB}^{medium}$ is reduced compared with their free-space value
$\sigma _{BB}^{free}$ by a factor of \cite{Persram02}
\begin{eqnarray}
R_{medium}(\rho,\delta,\vec{p})&\equiv& \sigma
_{BB}^{medium}/\sigma
_{BB}^{free}\nonumber\\
&=&(\mu _{BB}^{\ast }/\mu _{BB})^{2},
\end{eqnarray}
where $\mu _{BB}$ and $\mu _{BB}^{\ast }$ are the reduced masses
of the colliding baryon-pair in free space and medium,
respectively. This form of reduced elastic
baryon-baryon scattering cross section in medium agrees well with
our recent study \cite{yongp2}.
Since the inelastic baryon-baryon
scattering cross section in medium is less known but crucial for
$\pi$ production \cite{ko15}, we in the present model extend the
above reduced factor $R_{medium}(\rho,\delta,\vec{p})$ to
inelastic baryon-baryon scattering cross section \cite{yong2016}. Other treatments related to
$\pi$ production are similar to that in Ref. \cite{lyz05}.

For hard photon production from neutron-proton bremsstrahlung, we
use the prediction of the one boson exchange model by Gan et
al. \cite{gan94,yongp1,yongp2,yongp3}
\begin{equation}\label{QFT}
p_{\gamma}\equiv\frac{dN}{d\varepsilon_{\gamma}}=2.1\times10^{-6}\frac{(1-y^{2})^{\alpha}}{y},
\end{equation}
where $y = \varepsilon_{\gamma}/E_{max}$, $\alpha = 0.7319-0.5898\beta_i$,
$\varepsilon_{\gamma}$ is energy of emitting photon, $E_{max}$ is the energy
available in the center of mass of the colliding proton-neutron
pairs, $\beta_i$ is the initial velocity of
the proton in the proton-neutron center of mass frame.
The pauli-blockings of final state scattering neutron and proton in $pn\rightarrow pn\gamma$ process are taken into account \cite{bau86}.


\begin{figure}[h!]
\centering
\includegraphics[width=0.5\textwidth]{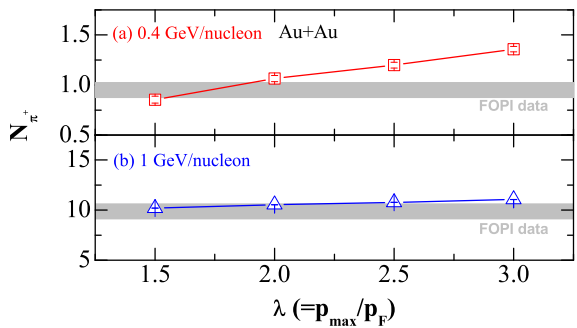}
\caption{ (Color online) The number of produced $\pi^{+}$ meson as a function of
high-momentum cutoff parameter $\lambda$ in the Au + Au collisions
at, respectively, 0.4 and 1 GeV/nucleon beam energies.}
\label{pion}
\end{figure}
Fig.~\ref{pion} shows $\pi^{+}$ production as a
function of high-momentum cutoff parameter $\lambda$ of colliding
nuclei in the Au + Au collisions at 0.4 and 1 GeV/nucleon
incident beam energies, respectively. One can clearly see that as the
high-momentum cutoff parameter $\lambda$ increases, more
$\pi^{+}$'s are produced. Larger high-momentum cutoff parameter
$\lambda$ causes larger nucleon average kinetic energy, especially
proton average kinetic energy \cite{sci14}, thus the average
center-of-mass energy of proton-proton collision also becomes
larger. As a consequence more $\pi^{+}$'s are produced in
nucleus-nucleus collision \cite{stock86}. This is the reason why
one sees in the upper panel of Fig.~\ref{pion} more $\pi^{+}$'s
are produced with large high-momentum cutoff parameter $\lambda$.
As incident beam energy increases, the initial movement of
nucleons in nuclei becomes less important in nucleus-nucleus
collisions. We thus see, in the lower panel of Fig.~\ref{pion}, at
1 GeV/nucleon incident beam energy, $\pi^{+}$ production is less
sensitive to the high-momentum cutoff parameter $\lambda$ (At 0.4 GeV/nucleon,
the sensitivity of $\pi^{+}$ production
to $\lambda$ is about 10 times that of $\pi^{+}$ at 1 GeV/nucleon). Fig.~\ref{pion}
shows $\lambda \leq 2$ is favored by the FOPI data \cite{npa2010}.

\begin{figure}[h!]
\centering
\includegraphics[width=0.5\textwidth]{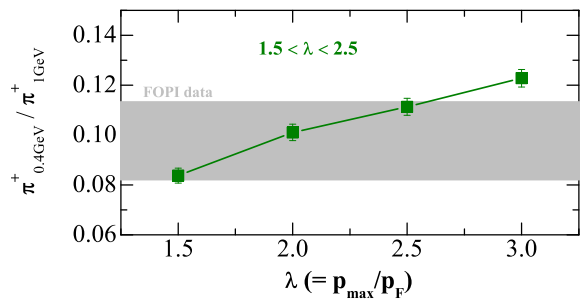}
\caption{ (Color online) Constraints on the high-momentum cutoff parameter
$\lambda$ by the ratio of $\pi^{+}$ productions in the Au + Au
collision at low and high beam energies. The shadow region denotes
the FOPI data \cite{npa2010}. } \label{const}
\end{figure}
Since $\pi^{+}$ production is sensitive to the high-momentum
cutoff parameter $\lambda$ only at relative low beam energy, one
can construct the ratio of $\pi^{+}$'s produced from low and high
incident beam energies to probe the high-momentum cutoff
parameter. This ratio is expected to reduce the system errors in some degree.
Shown in Fig.~\ref{const} is the ratio of $\pi^{+}$
productions at 0.4 and 1 GeV/nucleon incident beam energies as a
function of high-momentum cutoff parameter $\lambda$. As expected,
the ratio of $\pi^{+}$ multiplicities produced respectively from low
and high beam energies are still very sensitive to the
high-momentum cutoff parameter $\lambda$. By comparison with the
FOPI pion production data \cite{npa2010}, a constraint of
1.5 $\leq$ $\lambda$ $\leq$ 2.5 is obtained. This result is surprisingly similar to
that in Ref.~\cite{Egiyan2006}.

The hadronic probe $\pi^{+}$ production inevitably suffers from distortions due to the strong
interactions in the final state. Ideally one would like to have more clean ways to probe the high-momentum tail of nucleon in nucleus. Photons interact with nucleons only electromagnetically, once produced they escape almost freely from the nuclear environment
in nuclear reactions. In this regard, we also use hard photon production to constrain the high-momentum cutoff parameter $\lambda$.

Hard photon production in heavy-ion reactions at beam energies
below 200 MeV/nucleon had been in fact extensively studied both
experimentally and theoretically \cite{bertsch,grosse86,nif90,cassrp,yongp1,yongp2,yongp3}.
The TAPS collaboration carried out a series of
comprehensive measurements studying in detail the properties of
hard photons \cite{TAPS,TAPS2,TAPS3,TAPS4}. Theoretically, it was concluded that the
neutron-proton bremsstrahlungs in the early stage of the reaction
are the main source of high energy $\gamma$ rays \cite{may1,may2}.
And it was demonstrated that the hard photons can be used to
probe the reaction dynamics leading to the formation of dense
matter \cite{bertsch86,ko85,cassing86,bau86,stev86}. And
effects of the nuclear Equation of State (EOS) on the hard photon
production were found small \cite{ko87}. Although the input
elementary $pn\rightarrow pn\gamma$ probability is still model
dependent \cite{nif85,nak86,sch89,gan94,tim06}, the experimental data can be
described reasonably well theoretically within a factor of 2 \cite{cassrp}.
And the experimental efforts have the potential to improve the
situation significantly in the near future \cite{saf07}.

\begin{figure}[h!]
\centering
\includegraphics[width=0.5\textwidth]{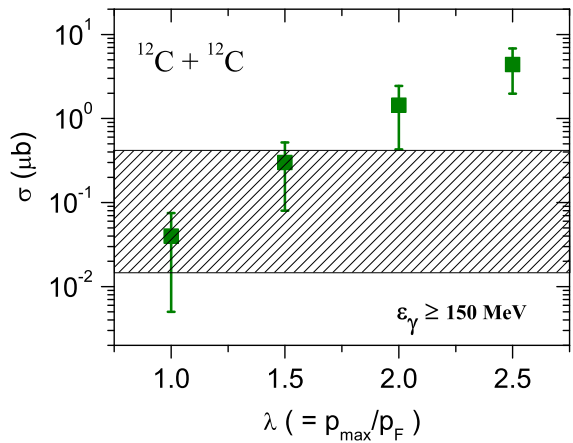}
\caption{ (Color online) Inclusive
photon production cross sections ($\varepsilon_{\gamma}\geq$ 150 MeV)
in $^{12}$C+$^{12}$C collisions at the beam energy of 60 MeV/nucleon.
The symbols stand for BUU calculations with, respectively, $\lambda$ = 1 (i.e., without HMT), 1.5, 2, 2.5. The shadow region denotes
experimental data \cite{yongp1,grosse86}. } \label{photo}
\end{figure}
Fig.~\ref{photo} shows comparison of theoretical inclusive
hard photon production cross sections in $^{12}$C+$^{12}$C collisions
and the experimental data \cite{yongp1,grosse86}. Since the hadronic probe $\pi^{+}$ production constrained
the high-momentum cutoff parameter $\lambda$ between 1.5 $\sim$ 2.5, we use
$\lambda$ = 1.5, 2, 2.5 as the BUU calculations with the high-momentum tail. As comparison, we also calculated the case without high-momentum tail ($\lambda$ = 1). From Fig.~\ref{photo},
it is seen that the hard photon production cross section in heavy-ion collisions
is very sensitive to the HMT of nuclei \cite{yongp3}. The BUU calculations with $\lambda$ = 2 and 2.5 are larger than
the experimental hard photon production cross section. And the case of BUU calculation without
HMT is somewhat lower than the experimental data. From Fig.~\ref{photo}, it is seen that
the electromagnetic probe hard photon constrains the
the high-momentum cutoff parameter to be $\lambda \leq$ 2. Combining the constraints from hadronic probe $\pi^{+}$ production (shown in Fig.~\ref{const}) and that from electromagnetic probe hard photon production (shown in Fig.~\ref{photo}), we can conservatively conclude that the value of the high-momentum cutoff parameter $\lambda$ in nuclei is less than 2.5 and the overlap-area is $\lambda \leq 2$, which is smaller than that deduced in other lecture \cite{hen14,henprc15,liba15}.

A small value of lambda implies lower average nucleon kinetic energy. The lower average nucleon kinetic energy implies smaller collision energy of nucleon pairs in transport calculations. This causes small number of meson production in heavy-ion collisions at low and intermediate energies. And it also cause smaller number of energetic nucleon or meson emissions, a high value of $\pi^{-}/\pi^{+}$ ratio \cite{zhang2016} and small number of hard photon production in heavy-ion collisions at low and intermediate energies. A small value of lambda also implies a relatively larger nuclear kinetic symmetry energy, thus causes the reduction of nuclear symmetry potential \cite{henprc15}. The reduction of nuclear symmetry potential in heavy-ion collisions at intermediate energies decreases the sensitivity of isospin-sensitive observables.

Because the high-momentum tail of nucleon momentum distribution is in fact caused by the short-range correlations of nucleons, while in our transport model (besides nucleon-nucleon or nucleon-meson collisions and nuclear pauli-blockings) only a mean-field potential is used. Thus the nuclei in the evolution before collision may be instable. Lacking of the binding of high-momentum nucleons in nuclei, the shape of initial distribution of nucleons in momentum space may be changed and energetic nucleons may escape out of the nuclei \cite{yong2016}. Therefore,
the average kinetic energy of nucleons in the reaction system decreases and then cause less $\pi^{+}$ meson or hard photon productions. The increased stability of colliding nuclei may cause somewhat more $\pi^{+}$ meson or hard photon productions (less emission of high-momentum nucleons corresponds to a higher average kinetic energy of nucleons in nuclei).
Thus considering stability of colliding nuclei, more pion and hard photon may be produced than that in the present work.
In a word, after considering stability of colliding nuclei and comparing to the same experimental data, the high-momentum cutoff parameter $\lambda$ should be smaller than our current constraint.
Also uncertainty of the mechanism of hard photon productions may affect the conclusion here.
The probability of hard photon production from the semiclassical hard
sphere collision model \cite{bertsch,nif90,cassrp} will give somewhat more photon production \cite{yongp1}, thus also require a smaller $\lambda$ value to explain the experimental data.
Furthermore, the off-shell transport of particle production in the BUU model may also cause more $\pi^{+}$ meson and hard photon productions \cite{ber96,cass00,mosel02}, thus also require a smaller $\lambda$ value to fit the experimental data. As in this work the $\lambda$ value is constrained to be $\lambda \leq 2$, while considering all the above factors, the $\lambda$ parameter should be not larger than 2.


In conclusions, based on the nuclear transport model, we studied how the
high-momentum cutoff parameter $\lambda$ affects $\pi$ and hard photon productions
in nucleus-nucleus collisions at intermediate energies. It is found
that $\pi^{+}$ and hard photon productions in nucleus-nucleus collision at lower
beam energy is very sensitive to the value of the high-momentum
cutoff parameter $\lambda$. By comparing the BUU's $\pi^{+}$ and hard photon
productions with experimental data and considering some uncertainties, a constraint of high-momentum cutoff value
$\lambda \leq 2$ is obtained.

Constraints on the high-momentum cutoff parameter
$\lambda$ in nuclei have implications in the studies of nuclear force
at short distance, in the construction of nuclear transport model of
heavy-ion collisions at intermediate energies, in the studies of
equation of state of dense nuclear matter and the nuclear symmetry
energy at suprasaturation densities or in the study of the physics in neutron stars, etc.


The work was carried out at National Supercomputer
Center in Tianjin, and the calculations were performed on
TianHe-1A. The work is supported by the National Natural Science
Foundation of China under Grant Nos. 11375239, 11435014.


\begin{thebibliography}{00}
\bibitem{e93} L. Lapikas, Nucl. Phys. A. {\bf 553}, 297 (1993).
\bibitem{e96} J. Kelly, Adv. Nucl. Phys. {\bf 23}, 75 (1996).
\bibitem{sci08} R. Subedi et al. (Hall A. Collaboration), Science {\bf 320}, 1476 (2008).
\bibitem{pia06} E. Piasetzky, M. Sargsian, L. Frankfurt, M. Strikman, J. W. Watson, Phys. Rev. Lett. {\bf 97}, 162504 (2006).
\bibitem{sh07} R. Shneor et al., Phys. Rev. Lett. {\bf 99}, 072501 (2007).
\bibitem{tenf05} M. M. Sargsian, T. V. Abrahamyan, M. I. Strikman and L. L. Frankfurt, Phys. Rev. C {\bf 71}, 044615 (2005).
\bibitem{tenf07} R. Schiavilla, R. B. Wiringa, S. C. Pieper and J. Carlson, Phys. Rev. Lett. {\bf 98}, 132501 (2007).
\bibitem{bethe71} H. A. Bethe, Ann. Rev. Nucl. Part. Sci. {\bf 21}, 93 (1971).
\bibitem{anto88} A. N. Antonov, P. E. Hodgson and I. Z. Petkov, \emph{Nucleon Momentum
and Density Distributions in Nuclei} (Clarendon Press, Oxford,
1988).
\bibitem{Rios09}A. Rios, A. Polls, and W. H. Dickhoff, Phys. Rev. C \textbf{79}, 064308 (2009).
\bibitem{yin13}P. Yin, J. Y. Li, P. Wang, and W. Zuo, Phys. Rev. C \textbf{87}, 014314 (2013).
\bibitem{Claudio15}Claudio Ciofi degli Atti, Physics Reports \textbf{590}, 1 (2015).

\bibitem{Ciofi96} C. Ciofi degli Atti, S. Simula, Phys. Rev. C {\bf 53}, 1689 (1996).
\bibitem{Fantoni84} S. Fantoni and V. R. Pandharipande, Nucl. Phys. A {\bf 427}, 473 (1984).
\bibitem{Pieper92} S. C. Pieper, R. B.Wiringa, and V. R. Pandharipande, Phys. Rev. C {\bf 46}, 1741 (1992).
\bibitem{egiyan03} K. Sh. Egiyan, et al., Phys. Rev. C {\bf 68}, 014313 (2003).
\bibitem{hen14} O. Hen, L. B. Weinstein, E. Piasetzky, G. A. Miller, M. M. Sargsian, and Y. Sagi, Phys. Rev. C {\bf 92}, 045205 (2015)
\bibitem{henprc15} O. Hen, B. A. Li, W. J. Guo, L. B. Weinstein, and E. Piasetzky, Phys. Rev. C {\bf 91}, 025803 (2015).
\bibitem{liba15} B. J. Cai, B. A. Li, Phys. Rev. C {\bf 92}, 011601 (2015).
\bibitem{sci14} O. Hen et al. (The CLAS Collaboration), Science {\bf 346}, 614 (2014).
\bibitem{Egiyan2006} K. S. Egiyan, et al., Phys. Rev. Lett. {\bf 96}, 082501 (2006).
\bibitem{Sargsian12} M. McGauley and Misak M. Sargsian, arXiv: 1102.3973v3 (2012).
\bibitem{wangp2013} P. Wang, S.-X. Gan, P. Yin and W. Zuo, Phys. Rev. C {\bf 87}, 014328 (2013).
\bibitem{zhang2016} F. Zhang, G. C. Yong,  arXiv: 1605.03656 (2016).

\bibitem{topic14} ``Topical issue on nuclear symmetry energy'', Eds., B. A. Li,
A. Ramos, G. Verde, and I. Vida\~{n}a, Eur. Phys. J. A {\bf 50},
No. 2, (2014).
\bibitem{Ryckebusch11}Jan Ryckebusch, Wim Cosyn, Maarten Vanhalst,
Phys. Rev. C {\bf 83}, 054601 (2011).
\bibitem{stock86} R. Stock, Phys. Rep., {\bf 135}, 259 (1986).
\bibitem{yongp1}G. C. Yong, B. A. Li, and L. W. Chen, Phys. Lett. B \textbf{661}, 82 (2008).
\bibitem{yongp2}G. C. Yong, W. Zuo, and X. C. Zhang, Phys. Lett. B \textbf{705}, 240 (2011).
\bibitem{yongp3}H. Xue, C. Xu, G. C. Yong, Z. Z. Ren, Phys. Lett. B \textbf{755}, 486 (2016).

\bibitem{bertsch} G. F. Bertsch and S. Das Gupta, Phys. Rep. {\bf 160}, 189 (1988).
\bibitem{Dan02a} P. Danielewicz, R. Lacey, W. G. Lynch, Science {\bf 298}, 1592 (2002).
\bibitem{Persram02} D. Persram and C. Gale, Phys. Rev. C {\bf 65}, 064611 (2002).

\bibitem{Sargsian14} Misak M. Sargsian, Phys. Rev. C {\bf 89}, 034305 (2014).
\bibitem{xu13} C. Xu, A. Li, B. A. Li, J. of Phys: Conference Series {\bf 420}, 012090 (2013).
\bibitem{Das03} C. B. Das, S. DasGupta, C. Gale, B. A. Li, Phys. Rev. C {\bf 67}, 034611 (2003).
\bibitem{xu14} J. Xu, L. W. Chen, B. A. Li, Phys. Rev. C {\bf 91}, 014611 (2015).
\bibitem{Isaac2011} Isaac Vidana, Artur Polls, Constanca Providencia, Phys. Rev. C {\bf 84}, 062801 (R) (2011).
\bibitem{Dieperink03} A. E. L. Dieperink, Y. Dewulf, D. VanNeck, M. Waroquier, V. Rodin, Phys. Rev. C {\bf 68}, 064307 (2003).
\bibitem{ko15} T. Song, and C. M. Ko, Phys. Rev. C {\bf 91}, 014901 (2015).
\bibitem{yong2016}G. C. Yong, Phys. Rev. C {\bf 93}, 044610 (2016).
\bibitem{lyz05} B. A. Li, G. C. Yong, W. Zuo, Phys. Rev. C {\bf 71}, 014608 (2005).
\bibitem{gan94} N. Gan et al., Phys. Rev. C \textbf{49}, 298 (1994).
\bibitem{bau86} W. Bauer, G.F. Bertsch, W. Cassing and U. Mosel, Phys. Rev. C
\textbf{34}, 2127 (1986).
\bibitem{npa2010} W. Reisdorf et al., Nucl. Phys. A {\bf 848}, 366 (2010).

\bibitem{grosse86} E. Grosse et al., Europhys. Lett. \textbf{2}, 9 (1986).
\bibitem{nif90} H. Nifenecker and J.A. Pinston, Annu. Rev. Nucl. Part. Sci., {\bf 40}, 113 (1990).
\bibitem{cassrp} W. Cassing, V. Metag, U. Mosel, and K. Niita,
Phys. Rep. \textbf{188}, 363 (1990).

\bibitem{TAPS} Y. Schutz et al. for the TAPS collaboration, Nucl. Phys. A {\bf 622}, 404
(1997).
\bibitem{TAPS2} G. Martinez et al., Phys. Lett. B {\bf 461}, 28 (1999).
\bibitem{TAPS3} David d'Enterria et  al., Phys. Lett. B {\bf 538}, 27 (2002).
\bibitem{TAPS4} R. Ortega et al., Eur. Phys. J. A {\bf 28}, 161 (2006).

\bibitem{may1}G. H. Liu et al., Phys. Lett. B {\bf 663}, 312 (2008).
\bibitem{may2}Y. G. Ma et al., Phys. Rev. C {\bf 85}, 024618 (2012).
\bibitem{bertsch86} B. A. Remington, M. Blann and G. F. Bertsch, Phys. Rev. Lett.
\textbf{57}, 2909 (1986).
\bibitem{ko85} C. M. Ko, G. F. Bertsch and J. Aichelin, Phys. Rev. C \textbf{31}, 2324(R) (1985).
\bibitem{cassing86} W. Cassing, T. Biro, U. Mosel, M. Tohyama, and W. Bauer, Phys. Lett. B
\textbf{181}, 217 (1986).

\bibitem{stev86} J. Stevenson et al., Phys. Rev. Lett. \textbf{57}, 555 (1986).
\bibitem{ko87} C. M. Ko and J. Aichelin, Phys. Rev. C {\bf 35}, 1976
(1987).
\bibitem{nif85} H. Nifenecker and J. P. Bondorf, Nucl. Phys. A \textbf{442}, 478 (1985).
\bibitem{nak86} K. Nakayama and G. F. Bertsch, Phys. Rev. C {\bf 34}, 2190 (1986).
\bibitem{sch89}M. Sch\"affer, T.S. Biro, W. cassing and U. Mosel,
H. Nifenecker and J.A. Pinstan, Z. Phys. A {\bf 339}, 391 (1991).
\bibitem{tim06} R. G. E. Timmermans, T. D. Penninga, B. F. Gibson, M. K. Liou, Phys. Rev. C \textbf{73}, 034006 (2006).
\bibitem{saf07} Y. Safkan et al., Phys. Rev. C \textbf{75}, 031001(R) (2007).

\bibitem{ber96} G. F. Bertsch, P. Danielewicz, Phys. Lett. B {\bf 367}, 55 (1996).
\bibitem{cass00} W. Cassing and S. Juchem, Nucl. Phys. A {\bf 665}, 377 (2000).
\bibitem{mosel02} A. B. Larionov and U. Mosel, Phys. Rev. C {\bf 66}, 034902 (2002).

\end{thebibliography}
\end{document}